\documentclass[prl,twocolumn,superscriptaddress]{revtex4}

\usepackage{epsfig}
\usepackage{amsmath}

\newcommand{\ran}{\rangle}
\newcommand{\lan}{\langle}

\newcommand{\order}{O}

\newcommand{\kb}[2]{ | #1 \rangle \langle #2 |}

\begin{document}

\title{Efficiently contractable quantum circuits cannot produce much entanglement}

\author{Nadav Yoran}\email{N.Yoran@bristol.ac.uk}
\affiliation{H.H.Wills Physics Laboratory, University of Bristol,
Tyndall Avenue, Bristol BS8 1TL, UK}


\begin{abstract}
We show a similarity between two different classical simulation
methods for measurement based quantum computation -- one relying on
a low entanglement (tree tensor network) representation of the
computer's state, and the other a tensor contraction method based on
the topology of the graph state. We use this similarity to show that
any quantum circuit that can be efficiently simulated via tensor
contraction cannot produce much entanglement.
\end{abstract}

\maketitle

Characterizing classes of quantum computations that can be simulated
efficiently on a classical computer is one of the most useful ways
for investigating the power of quantum computation. In particular
classical simulations can be used to identify the crucial properties
which are responsible for the advantage that quantum computation has
over classical computation. One property that has been shown as
essential for a valuable quantum computation is entanglement
(measured by the Schmidt number). Any computation that does not
produce much entanglement, in the sense that the state of the
computer, at any stage of the computation, can be described through
a sequence of Schmidt decompositions for which the Schmidt number is
bounded by poly$(n)$  ($n$ -- the number of qubits), can be
efficiently simulated on a classical computer \cite{vidal2,jozsa}.

One method for simulating quantum circuits which does not make use
of a low-entanglement representation of the computer's state is
tensor contraction due Markov and Shi \cite{markov}. This method
relies solely on the topology of the circuit, or more accurately, of
the graph created by representing the gates as vertices and the
qubit wires as edges. It was shown that a number of families of
quantum circuits including log-depth circuits with nearest neighbor
interaction \cite{markov} and the approximate quantum Fourier
transform (which was not known to be efficiently simulable by any
other method), can be efficiently simulated by this method
\cite{us,aharonov}. The relation between low-entanglement based
methods and tensor contraction was not completely clear. In
particular, the question how much entanglement can circuits whose
topology allows for an efficient classical simulation produce was
until now open.

In this work we first consider classical simulations of measurement
based quantum computation (MQC). We show a similarity between the
tree tensor network (TTN) simulation method of \cite{vidal,briegel}
-- which is a low-entanglement method based on a sequence of Schmidt
decompositions (relying on low enough Schmidt numbers) -- and a
version of the tensor contraction method, as defined in \cite{us},
for MQC. We then use this similarity to show that any quantum
circuit which is efficiently contractable cannot produce much
entanglement. Namely, during the entire computation the state of the
computer has an efficient representation (as a TTN) in terms of as
sequence of Schmidt decompositions.

In the methods above (as in all other simulation methods of quantum
computation that we know of) one simulates a quantum computation
through sampling in a qubit by qubit manner. Explicitly, one
computes the probabilities for the output measurement on one qubit
and samples from them, then one computes the conditional
probabilities for a measurement on a second qubit given that the
sampled outcome for the first qubit had been obtained and so on. At
each stage one computes the conditional probabilities for a
measurement on one qubit given that previously sampled outcomes had
been obtained. At the end of the process one obtains an outcome, of
the entire computation, with the same probability as the quantum
computer.

\begin{figure}\begin{center}
\epsfig{file=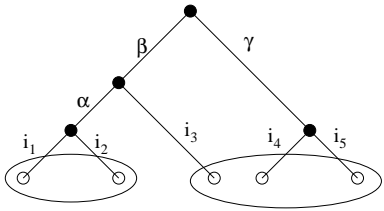}\caption{ A tree structure on a $5$ qubits
state. The partition corresponding to the edge $\alpha$ is shown.
\label{fig:tree5}}
\end{center}\end{figure}

Let us first consider the simulation method based on the
representation of the system as a TTN. This representation of a
quantum state is a generalization of Vidal's \cite{vidal2} matrix
product state (MPS) representation. In the MPS representation of a
system of $n$ qubits one considers $n-1$ Schmidt decompositions
according to a chosen ordering (that is, one considers the first
qubit against the rest then the first two qubit against the rest and
so on), whereas in the TTN representation one considers a sequence
of Schmidt decompositions along a chosen tree structure. Explicitly,
given an $n$ qubit system we consider a tree which has these qubits
as its leaves as in Fig. \ref{fig:tree5}. Removing an edge from the
tree will induce a partition of the tree into two sub-trees. The
Schmidt decompositions we consider in a TTN are those corresponding
to partitions induced by removing one internal edge of such a tree
graph. In \cite{vidal, briegel} a subcubic tree was considered where
all vertices, except the leaves, including the root have three
edges. Here we shall turn this into a binary tree by adding a degree
two vertex proportional to the identity (which will be the new root)
to one of the edges connected to the original root. In a TTN
representation a tensor is associated with each vertex of the tree.
The indices of these tensors are associated with the edges connected
to the vertex. Therefore in our TTN every tensor (except the root)
would be of rank $3$. Indices associated with edges connected to the
leaves correspond to the value of the qubit in the computational
basis. So that given an $n$ qubit state and a binary tree structure
on that state, the coefficients of the computational basis expansion
of the state are given in terms of a contraction over these tensors.
For example, a TTN representation of the state depicted in Fig.
\ref{fig:tree5} would be
\begin{equation}
 \sum_{i_1 \cdots i_5} \sum_{\alpha,\beta,\gamma}
 A_{i_1,i_2,\alpha}A_{i_3,\alpha,\beta}A_{\beta,\gamma}A_{i_4,i_5,\gamma}
 \, |i_1 \cdots i_5\ran
\end{equation}
where the $i$ indices are the value of the qubit (in the
computational basis) and the Greek indices correspond to the
internal edges of the tree and thus correspond to the Schmidt basis
of the partition induced by this edge. So that, for instance, the
Schmidt vectors corresponding to the decomposition along the edge
$\alpha$ (up to normalization) would be
\begin{eqnarray}
 |\Phi^{[1,2]}_{\alpha}\ran = \sum_{i_1,i_2}
 A_{i_1,i_2,\alpha}|i_1,i_2\ran \; , \nonumber \\
 |\Phi^{[3,4,5]}_{\alpha}\ran = \sum_{i_3,i_4,i_5,\beta,\gamma}
 A_{i_3,\alpha,\beta}A_{\beta,\gamma}A_{i_4,i_5,\gamma}|i_3,i_4,i_5\ran
 \: .
\end{eqnarray}
Denoting the largest Schmidt number for a certain tree by $\chi$,
the number of complex parameters describing the state is of the
order $O(n\chi^{3})$.

It was shown \cite{vidal} that using a TTN representation the
response of the the system to local operations and classical
communication (LOCC) as well as two-qubit operations can be
efficiently simulated classically. Therefore, one can efficiently
simulate any computation that does not generate much entanglement
(where $\chi$ is bounded by poly$(n)$). Furthermore, since in MQC
the computation is performed by LOCC that can not increase the
entanglement, any MQC can be efficiently simulated given that the
initial state has an efficient TTN representation.

Let us now consider the tensor contraction method for simulating
quantum circuits \cite{markov,us}. In this method we first associate
a graph with the circuit, representing each input qubit, gate and
output measurement by a vertex, and each wire by an edge (e.g. a two
qubit gate would correspond to a vertex of degree four). We then
label each edge with a different index. Finally, to each vertex we
associate a tensor describing the operation performed at that point.
This tensor has indices corresponding to all edges connected to that
vertex (so that its rank is equal to the degree of the vertex). The
probability to obtain a certain outcome for some measurement is
given by contracting these tensors, (that is, taking the product of
all these tensors and summing over all indices).

Of course, one cannot efficiently sum over all indices at the same
time as there are exponentially many terms. To avoid this one
contracts the tensors together one at a time -- in each step of the
computation one replaces two existing tensors with a new one
obtained by summing over any joint indices. We repeat this procedure
until we are left with a single tensor with no free indices, which
is the desired probability. The key element of the method is the
ordering of the contraction. The aim is to order the contractions so
that one never generate tensors with too many indices during this
process.

The contraction process can be described \cite{us} by a sequence of
sets of vertices $\cal{S}$ $=(s^{1},\ldots,s^{N})$ - each of which
corresponds to a particular tensor that is generated during the
computation. Each set $s^i \in \cal{S}$ is either the union of two
previous sets, or one previous set and a vertex, or two vertices.
Denoting the set of all vertices by $V$:
\begin{equation}
 s^{i}=\{t_{1}^{i}\cup t_{2}^{i}\} \quad \mbox{where} \quad \begin{array}{l}
 \mbox{either}\;\;\: t_{j}^{i}=s^{k}\: , \: k<i \: , \\ \mbox{or}\quad\quad \:
 t_{j}^{i}=\{v\} \: , \:v \in V. \end{array}
\end{equation}
The calculation of the probability is done in $N$ steps, where in
step $i$ we compute a new tensor by summing over all indices
corresponding to edges connecting $t_{1}^{i}$ to $t_{2}^{i}$. For
the computation to be complete, we require that the final set
$s^{N}=V$. The number of edges connecting vertices in $s^{i}$ to
vertices outside $s^{i}$ ($L^{i}$) is the rank of the tensor
corresponding to $s^{i}$. The simulation corresponding to the
sequence $\cal S$ will be an efficient one if $L^{max}=\max_i
L^{i}=\order(\log n)$.

In what follows we consider simulations of MQC. The state used as
the resource for MQC is the graph state, which we consider here as
any state that is generated by taking a set of qubits with each of
them initially in the state $|+\ran$ and applying cphase operations
to any number of pairs of qubits (the order in which these
operations are applied is irrelevant, one would always end up in the
same state) we will refer to these operations as cphase
connections). The underlying graph ($G$) of such a graph state is
obtained by associating a vertex to each qubit and an edge to each
cphase connection.

A natural way of defining a tensor contraction scheme for MQC is by
using PEPS \cite{verstraete}. In such a scheme a tensor is
associated with each of the qubits in the graph state. Such a tensor
corresponding to a particular qubit would have an index for each
cphase connection plus an additional index for the value of the
qubit. The PEPS representation of a graph state would be of the form
\begin{equation}
 |\Psi \ran = \sum_{i_{1}\cdots i_{n}} \sum_{ \{ \alpha \}}
 A^{i_{1}}_{ \{ \alpha \}_{1}} \cdots A^{i_{n}}_{ \{ \alpha \}_{n}} |i_{1}
 \cdots i_{n} \ran \: ,
 \label{peps}
\end{equation}
where $\{\alpha\}$ donates the set of all the indices corresponding
to cphase connections in the graph, $\{\alpha\}_{j}$ denotes the
subset of indices corresponding to cphase connections of qubit $j$.

In order to simulate MQC on a graph state it is enough to calculate
the probabilities for single qubit measurements. In more general
terms, we consider a measurement described by a set of POVM elements
$\{E_{r}\}$ each composed of single qubit POVM elements. Namely,
$E_{r}=E_{r_1}\otimes \cdots \otimes E_{r_n}$ where $r_{j}$
represents an outcome of a single qubit measurement. For a
projective measurement (on qubit $j$) $E_{r_{j}}=\kb{r_{j}}{r_{j}}$,
and for an unmeasured qubit we define $E_{r_{j}}:=I$. The
probability ($P(r)$) for obtaining a certain outcome $r=r_1\cdots
r_n$ is given by $tr(E_{r}\, \kb{\psi}{\psi})$, and from
\eqref{peps} we have.
\begin{equation}
 P(r)= \sum_{ \{\alpha \}\{ \alpha\}^{\prime}}
 B^{[1]r_{1}}_{ \{\alpha\}_{1} \{ \alpha\}_{1^{\prime}}}
 B^{[2]r_{2}}_{ \{\alpha\}_{2} \{ \alpha\}_{2^{\prime}}}
 \cdots B^{[n]r_{n}}_{ \{\alpha\}_{n} \{ \alpha\}_{n^{\prime}}}
 \label{prob}
\end{equation}
where
$$
 B^{[j]r_{j}}_{ \{\alpha\}_{j} \{ \alpha \}_{j^{\prime}} } = \sum_{i_{j}i^{\prime}_{j}}
 A^{i_{j}}_{ \{\alpha\}_{j} } A^{*\,j^{\prime}_{j}}_{ \{\alpha \}
 _{j^{\prime}} } \langle i^{\prime}_{j}| E_{r_j}| i_{j}\ran
$$
The probability is, therefore, given in terms of a tensor
contraction scheme on the same graph as the underlying $G$, where a
tensor $B^{[k]}$ is associated with the vertex corresponding to
qubit $k$. Note that in this tensor network there is a pair of joint
indices for each edge in the graph -- one from the subset
$\{\alpha\}$ and the corresponding one from $\{\alpha\}^{\prime}$ --
we can treat the pair as a single index admitting four values. The
above expression \eqref{prob} for the probability can be obtained by
placing the graph associated with the bra $\lan \psi|$ and its
tensor network on 'top' of the graph of $|\psi \ran$ (the two graphs
are similar only the associated tensor network of the first is the
complex conjugate of the other) and summing over the indices of the
qubits (the $i_{k}$'s) with the appropriate measurement elements in
between.

Once we have expressed the probabilities in terms of a contraction
scheme over a graph (namely, the probability is given by contracting
a set tensors where each tensor is associated with a vertex and each
index of these tensors is associated with an edge) then all the
results obtained in \cite{us} and \cite{markov} automatically apply
here.

The key point here is that a sequence $\cal{S}$ defined on a system
of qubits is completely equivalent to a binary tree structure $T$ on
the same system -- $\cal{S}$ defines a tree $T$ and vice versa. Any
subset $s^{k} \in \cal{S}$ of vertices (which now corresponds to
some set of qubits) corresponds to a subtree of $T$.

What can we learn from the above equivalence? A bipartition of the
system corresponding to an internal edge $e$ in the tree $T$ (that
is the partition induced by removing $e$), is simply a partition to
a set in $\cal{S}$ and the rest of the system. Let us denote the
Schmidt number corresponding to this bipartition by $\chi_{e}$ and
the number of cphase-connections between qubits corresponding to
vertices in the set and the rest of the system (or in other words
the number of edges in $G$ connecting vertices in the set to
vertices outside the set) by $L_{e}$. Clearly $L_{e} \geq \log_{2}
\chi_{e}$. As this is true for any $e \in T$ it is certainly true
for the one with the maximal Schmidt number. The edge in $T$ for
which we obtain the maximal $L_{e}$ might be a different one but
obviously the corresponding partition cannot have less
cphase-connections. Thus, for any graph state $|\psi\ran$ and any
tree structure on it $T$ we have
\begin{equation}
 \mathop{\mbox{max} }\limits_{e \in T} L_{e}(|\psi \ran ,T) \geq \mathop{\mbox{max} }
 \limits_{e^{\prime} \in T} \log_{2}\chi_{e^{\prime}}(|\psi \ran, T) \label{max}
\end{equation}
The minimum of the left hand side over all possible trees is the
contraction complexity of the underlying graph $G$ denoted by cc(G),
which determines the complexity of the best tensor contraction
scheme on the graph state. Classically simulating any computation
performed on $|\psi\ran$ would require $\mbox{poly}(n,2^{cc(G)})$
computational resources \cite{markov}. The minimum of the right hand
side of \eqref{max} over all possible trees is the Schmidt-rank
width (or $\chi$-width) of the graph state --
$\chi_{wd}(|\psi\ran)$, which determines the size of the most
efficient TTN description of the graph state (requiring ${\cal O}(n
2^{3\chi_{wd}})$ complex parameters). Classical simulation of any
computation performed on this state would require
$\mbox{poly}(n,2^{\chi_{wd}})$ computational resources
\cite{vidal,briegel}. As \eqref{max} applies to any tree structure
$T$ it applies also to the particular tree for which the minimum of
the left hand side is obtained. The minimum of the right hand side
therefore cannot be bigger. Thus
\begin{equation}
 cc(G) \geq \chi_{wd}(| \psi\ran) \label{chiwd}
\end{equation}
Hence, if we have an efficient tensor contraction simulation of MQC
then we are assured that there is an efficient simulation of the
same computation in terms of a TTN. Note that the opposite claim is
not true. Indeed, the number of cphase-connections between two parts
of a graph state gives an upper bound to $\log_{2}\chi_{e}$, however
there can be graph states where the number of connections greatly
exceeds $\log_{2}\chi_{e}$. The fully connected graph, for example,
where each of the qubits is connected to all the rest, has Schmidt
number $2$ for any possible bipartition of the state.

Yet, the graph states that one usually considers as a resource for
quantum computation are those without much excess of cphase
connections. In particular, graph states where the number of
connections per qubit (the degree of the vertices) is bounded by
some constant -- $\Delta$. In that case, for a bipartition of the
system corresponding to edge $e$, into a subtree (A) and the rest of
the system (B), the maximal number of qubits in A connected to
qubits in B is $\Delta\log_{2}\chi_{e}$. We can verify this by the
following procedure.
We construct a sequence of sets of qubits ($F_1, F_2, \cdots$).
$F_1$ includes a single qubit in $B$ and all the qubits that are
connected to it in $A$. $F_2$ consists of a different qubit in $B$
and all the qubits that are connected to it in $A$, except those
that are already included in $F_1$ and so on. We keep constructing
such sets until each of the qubits in $A$ that are connected to
qubits in $B$ is included in one of those sets. We now undo all
cphase connections within $A$ and $B$ (not affecting the
entanglement between the two sides). Next we measure (in the $z$
basis) all the qubits in $F_{1} \cap A$ save one, so that we are
left with a maximally entangled pair, where the qubit in $F_1 \cap
B$ is connected only to its counterpart in $F_1 \cap A$. The qubit
in $F_1 \cap A$ might be connected also to other qubits in $B$
however these connections can be undone by {\it local
complementation} \cite{hein} (i.e. by local clifford operations) and
unitary operation local to $B$. Thus, by using operations which
could only decrease the entanglement we have produced out of the
first set a maximally entangled pair with no connections to other
qubits. At that stage the qubit in $F_2 \cap B$ is connected only to
qubits within $F_2$, therefore we can repeat the procedure and
produce a maximally entangled pair also from this set. We proceed in
the same way generating a maximally entangled pair from each set,
each carrying exactly one ebit. Clearly, the maximal number of
pairs, and therefore of sets, cannot be greater than
$\log_{2}\chi_{e}$. As the maximal number of qubits from $A$ in each
set is $\Delta$ we get the above bound. The upper bound on $L_e$ is
thus $\Delta^{2}\log_{2}\chi_{A,B}$ since each qubit in A is
connected to at most $\Delta$ qubits in B. Since this bound applies
to any bipartition it also applies to the one for which $L_{e}$ is
maximal for a given tree. Obviously, the maximal $\chi_e$ for this
tree can only be greater, thus
\begin{equation}
 \mathop{\mbox{max} }\limits_{e \in T} L_{e}(|\psi \ran ,T) \leq \Delta^{2}\mathop{\mbox{max} }
 \limits_{e^{\prime} \in T} \log_{2}\chi_{e^{\prime}}(|\psi\ran,T) \label{min}
\end{equation}
Using the same arguments as above we can take a minimum over all
trees for both sides of the inequality. Including \eqref{max} we
therefore have
\begin{equation}
 \chi_{wd}(|\psi \ran) \leq cc(G) \leq \Delta^{2}\chi_{wd}(|\psi \ran) \label{min2}
\end{equation}
Showing the equivalence, up to a constant, of both methods of
simulations of MQC for any $G$ of maximal degree $\Delta$. A
topological parameter which determines $cc(G)$ is the {\it tree
width} of $G$ (twd($G$)). In \cite{markov} upper and lower bounds to
$cc(G)$ were given in terms of twd($G$). $\chi_{wd}$(G) is equal to
a different parameter of the graph -- the {\it rank width} of G
(rwd(G)) \cite{oum,briegel}. Using the bounds in \cite{markov} the
above inequalities can be written in terms of these parameters. For
any graph $G$ we have
$$
\frac{ \mbox{twd(G)}-1}{2\Delta^{2}}\, \leq \, \mbox{rwd(G)}\, \leq
\, \Delta(\mbox{twd(G)}+1) -1
$$

So far we have considered simulation of quantum computation on graph
states. Let us now consider quantum circuits. Given a quantum
circuit ($C$) we can on one hand define a a tensor contraction
scheme as in \cite{us,markov,aharonov} where the corresponding graph
($G_{c}$) is constructed by associating a vertex to each circuit
element. On the other hand we can perform the same computation using
MQC on a graph state. The standard way to construct a graph state
for a given circuit is by associating a chain of qubits for each
logical qubit along which the data would progress via single qubit
measurements while undergoing one and two qubit gate applications.
For each single qubit gate up to three qubits should be introduced
to the chain and two qubit gates are realized by introducing cphase
connections between the two corresponding chains. We denote, as
before, the underlying graph of the graph state by $G$.

Clearly $G_{c}$ is not identical to $G$. One difference is that in
$G$ a sequence of up to three vertices of degree $2$ may correspond
to a single one-qubit gate (which corresponds to one such vertex in
$G_{c}$). The second difference is that a two qubit gate is
associated with a single vertex of degree $4$ while in $G$ it would
correspond to two vertices of degree $3$ with an edge between them.
Clearly, the additional vertices of degree $2$ have no effect on the
contraction complexity of $G$ -- any degree $2$ vertex can be
contracted together with a neighboring vertex or a set of vertices
without changing its number of outgoing edges.
Since the two vertices of degree $3$ can be combined to form a set
with four outgoing edges (representing a rank $4$ tensor just as the
corresponding vertex in $G_{c}$), it is clear that
$$
 cc(G) \leq cc(G_{c})
$$
Hence, if a quantum circuit has an efficient tensor contraction
scheme simulating it then we are assured that so would the MQC
version of that circuit on a graph state, and from \eqref{chiwd} we
know that the this graph state would have an efficient TTN
representation and consequently an efficient TTN-based classical
simulation. Moreover, the fact that any such circuit has an
efficient TTN representation also tells us that such a circuit
cannot produce much entanglement. The state of the computer at any
stage of the computation would have an efficient TTN representation.

In order to see this we note that the state of the computer at a
certain stage of the computation is the state produced by applying a
sub-circuit of $C$ ($C_{1}$) to the input. In order to produce the
same state using our graph state we introduce new 'output' qubits in
the graph state corresponding to $C$. These are additional qubits
inserted in the chains just after the sub-graph state corresponding
to $C_{1}$, which function as links connecting this sub-graph to the
rest of the system (they have no vertical cphase connections between
different chains). The reason we introduced these new qubits is that
the qubits that immediately follow $C_{1}$ might have additional
connection between the chains. Obviously the vertices corresponding
to the new output qubits are of degree $2$ so that if the original
graph had an efficient TTN representation so would the new graph
with the additional qubits. Now, the state of the quantum computer
after applying $C_{1}$ would be the state (up to local phase
corrections) of the new output qubits after measuring the qubits
corresponding to $C_{1}$ according to the circuit, and measuring the
rest of the qubits in the $z$ basis. As single qubit measurements
cannot increase the Schmidt rank the new output qubits would also
have an efficient TTN representation.

It should be noted that the opposite claim is not true. There are
circuits which do not produce much entanglement but do not have an
efficient tensor contraction scheme. Obvious examples are classical
circuits (with 'classical' computational basis input and output
measurements) such as modular exponentiation, which do not produce
any entanglement and yet are not likely to have an efficient tensor
contraction scheme \cite{us2}.

\acknowledgments

The author wish to thank S. Popescu, A. J. Short, R. Jozsa, P.
Skrzypczyk, H. Briegel, M. Van den Nest and W. D\"{u}r for fruitful
discussions. This work was supported by UK EPSRC grant
(GR/527405/01), and by the UK EPSRC's ``QIP IRC'' project.




\begin{thebibliography}{99}

\bibitem{vidal2} G. Vidal, Phys. Rev. Lett. \textbf{91}, 14902
(2003).

\bibitem{jozsa} R. Jozsa, N. Linden, Proc. Roy. Soc. A
\textbf{459}, 2011 (2003).

\bibitem{markov} I. Markov and Y. Shi, quant-ph/0511069.

\bibitem{us} N. Yoran and A. J. Short, Phys. Rev. A \textbf{76}, 042321 (2007).


\bibitem{aharonov} D. Aharonov, Z. Landau and J. Makowsky, quant-ph/0611156.

\bibitem{briegel} M. Van den Nest, W. D\"{u}r, G. Vidal, H. J. Briegel, Phys. Rev. A \textbf{75},
 012337 (2007).

\bibitem{vidal} Y.Shi, L. Duan, and G. Vidal, Phys. Rev. A \textbf{74}, 022320
 (2006).

\bibitem{verstraete}  F. Verstraete, J. I. Cirac, cond-mat/0407066.


\bibitem{hein} M. Hein, W. Dur, J. Eisert, R. Raussendorf, M. Van den
Nest, proceedings of the International school of Physics "Enrico
Fermi" on "Quantum Computers, Algorithms and Chaos",  Varenna,
Italy, July, 2005.

\bibitem{oum} S.-I. Oum, Ph.D. thesis, Princeton University, 2005.

\bibitem{us2} N. Yoran and A. J. Short, Phys. Rev. A \textbf{76}, 060302(R) (2007).





\end{thebibliography}
\end{document}